\documentclass[runningheads]{llncs}
\usepackage{graphicx}
\usepackage{amsmath,amssymb,amsfonts}
\usepackage{flushend}
\usepackage{algorithmic}
\usepackage{cleveref}
\usepackage{mdframed,lipsum}
\usepackage{textcomp}
\usepackage{amsmath}
\usepackage{amssymb}
\usepackage[caption=false]{subfig}
\usepackage{booktabs}
\usepackage{xcolor}
\usepackage{tabularx}
\usepackage{booktabs}
\usepackage{multirow}
\usepackage{enumitem}
\usepackage{url}
\usepackage{lipsum}

\makeatletter
  \newcommand\tablescript{\@setfontsize\tablescript{8pt}{6}}
\makeatother

\usepackage{xspace}

\newcommand{\ie}{\textit{i.e.,}\xspace}
\newcommand{\eg}{\textit{e.g.,}\xspace}

\newcommand{\etal}{\textit{et al.}\xspace}

\newcommand{\tdd}{TDD\xspace}
\newcommand{\yw}{YW\xspace}
\newcommand{\bsk}{BSK\xspace}
\newcommand{\mra}{MRA\xspace}

\newcommand{\str}{\texttt{STR}\xspace}
\newcommand{\pls}{PLS\xspace}
\newcommand{\ars}{ARS\xspace}
\newcommand{\dom}{DOM\xspace}
\newcommand{\lik}{LIK\xspace}

\newcommand{\apppls}{\texttt{APP\textsubscript{\pls}}\xspace}
\newcommand{\appars}{\texttt{APP\textsubscript{\ars}}\xspace}
\newcommand{\appdom}{\texttt{APP\textsubscript{\dom}}\xspace}
\newcommand{\applik}{\texttt{APP\textsubscript{\lik}}\xspace}

\newcommand{\imppls}{\texttt{IMP\textsubscript{\pls}}\xspace}
\newcommand{\impars}{\texttt{IMP\textsubscript{\ars}}\xspace}
\newcommand{\impdom}{\texttt{IMP\textsubscript{\dom}}\xspace}
\newcommand{\implik}{\texttt{IMP\textsubscript{\lik}}\xspace}

\newcommand{\tespls}{\texttt{TES\textsubscript{\pls}}\xspace}
\newcommand{\tesars}{\texttt{TES\textsubscript{\ars}}\xspace}
\newcommand{\tesdom}{\texttt{TES\textsubscript{\dom}}\xspace}
\newcommand{\teslik}{\texttt{TES\textsubscript{\lik}}\xspace} 

\begin{document}
\title{An Empirical Assessment on  Affective Reactions of Novice Developers when Applying Test-Driven Development}
\titlerunning{An Empirical Assessment on  Affective Reactions of Novice Developers}
%
\author{Simone Romano\inst{1} \and
Davide Fucci\inst{2} \and Maria Teresa Baldassarre\inst{1} \and Danilo Caivano\inst{1} \and Giuseppe Scanniello\inst{3}}

\authorrunning{S. Romano et al.}
%
\institute{University of Bari, Bari, Italy \\ \email{\{simone.romano,mariateresa.baldassarre,danilo.caivano\}@uniba.it}\\
\and
University of Hamburg, Hamburg, Germany\\
\email{fucci@informatik.uni-hamburg.de}
\and
University of Basilicata, Potenza, Italy\\
\email{giuseppe.scanniello@unibas.it}}
\maketitle              
\begin{abstract}
We study whether and in which phase Test-Driven Development (\tdd) influences  affective states of novice developers in terms of pleasure, arousal, dominance, and liking. We performed a controlled experiment with 29 novice developers. Developers in the treatment group performed a development task using TDD, whereas those in the control group used a non-TDD development approach. We compared the affective reactions to the development approaches, as well as to the implementation and testing phases, exploiting a lightweight, powerful, and widely used tool, \ie Self-Assessment Manikin. We observed that there is a difference between the two development approaches in terms of affective reactions. Therefore, it seems that affective reactions play an important role when applying \tdd and their investigation could help researchers to better understand such a development approach. 


\end{abstract}
\section{Introduction}\label{sec:intro}
Test-Driven Development (\tdd) is an Agile software development approach in which a developer first writes a unit test to frame a chunk of functionality and then writes production code to make the test pass and applies refactorings to improve the internal quality of production and test code.
This iterative process happens in fast-paced iterations of five to ten minutes~\cite{Bec03}. 

TDD promises to increase external quality of software (\ie less functional bugs) and developers' productivity as: \textit{(i)} writing test first forces developers to break a problem into simpler ones; \textit{(ii)} the tests provide initial software quality assurance; and \textit{(iii)} the regression test suite resulting after several iterations allows the developer to catch breaking changes early. The safety net provided by the regression tests boosts developers' confidence to the extent that \tdd is referred to as ``The  art of fearless programming''~\cite{JM07}. 
However, empirical research on the effects of \tdd has so far shown inconclusive results \cite{MMP14,RM13,BLD10}. 
Some research relates these results to the negative affective states that developers experience when initially exposed to \tdd---\eg frustration due to the counter-intuitive behavior of designing test cases rather than immediately working on a~solution~\cite{BLD10}.

Recent studies have leveraged affective states of developers to improve requirements engineering~\cite{CHG10}, software development~\cite{GFW18}, and software evolution~\cite{OMD16}.
Further, sentiment analysis has been applied to study the collaborative facets of software development~\cite{GLN17}.
These previous studies are based on the analysis of artifacts, mostly in textual form, produced during the software development life-cycle. Graziotin~\etal~\cite{GFW18} showed that unhappiness (\ie experiencing a sequence of negative affective states) impacts developers' productivity. 

Although there is a growing interest in studying the affective states of developers and previous research hypothesizes that TDD elicits negative and positive affects (\eg counter-intuitive order and regression tests), no work has investigated whether and in which phase \tdd  influences affective states of (novice) developers. To fill this gap, we conducted a controlled experiment with 29 novice developers. Our experimental design allowed us to isolate the affective reactions to \tdd from a baseline---\ie ``Your Way development'' (\yw)---in terms of four dimensions: pleasure, arousal, dominance, and liking. To measure these dimensions, we relied on a lightweight yet powerful tool, namely Self-Assessment Manikin (SAM)~\cite{Bradley:1994}. 

The results of our study provide initial evidence that novice developers like \tdd less than \yw. Moreover, developers following \tdd seem to like the implementation phase less than the others, and the testing phase seems to make developers using \tdd less happy. To foster replications of our study so increasing the confidence in this initial evidence, we make our laboratory package public.\footnote{\url{https://bit.ly/2UGn4o1}}

\textbf{Paper Structure.} 
 Section~\ref{sec:relatedWork} discusses background  and  related work. Section~\ref{sec:planning} details the planning of our experiment. The results from the experiment are presented in Section~\ref{sec:results} and discussed in Section~\ref{sec:discussion}. Possible limitations are reported in Section~\ref{sec:threats}. 
Section~\ref{sec:conclusion} concludes the paper.

\begin{figure}[t]
    \centering
    \includegraphics[width=0.65\columnwidth]{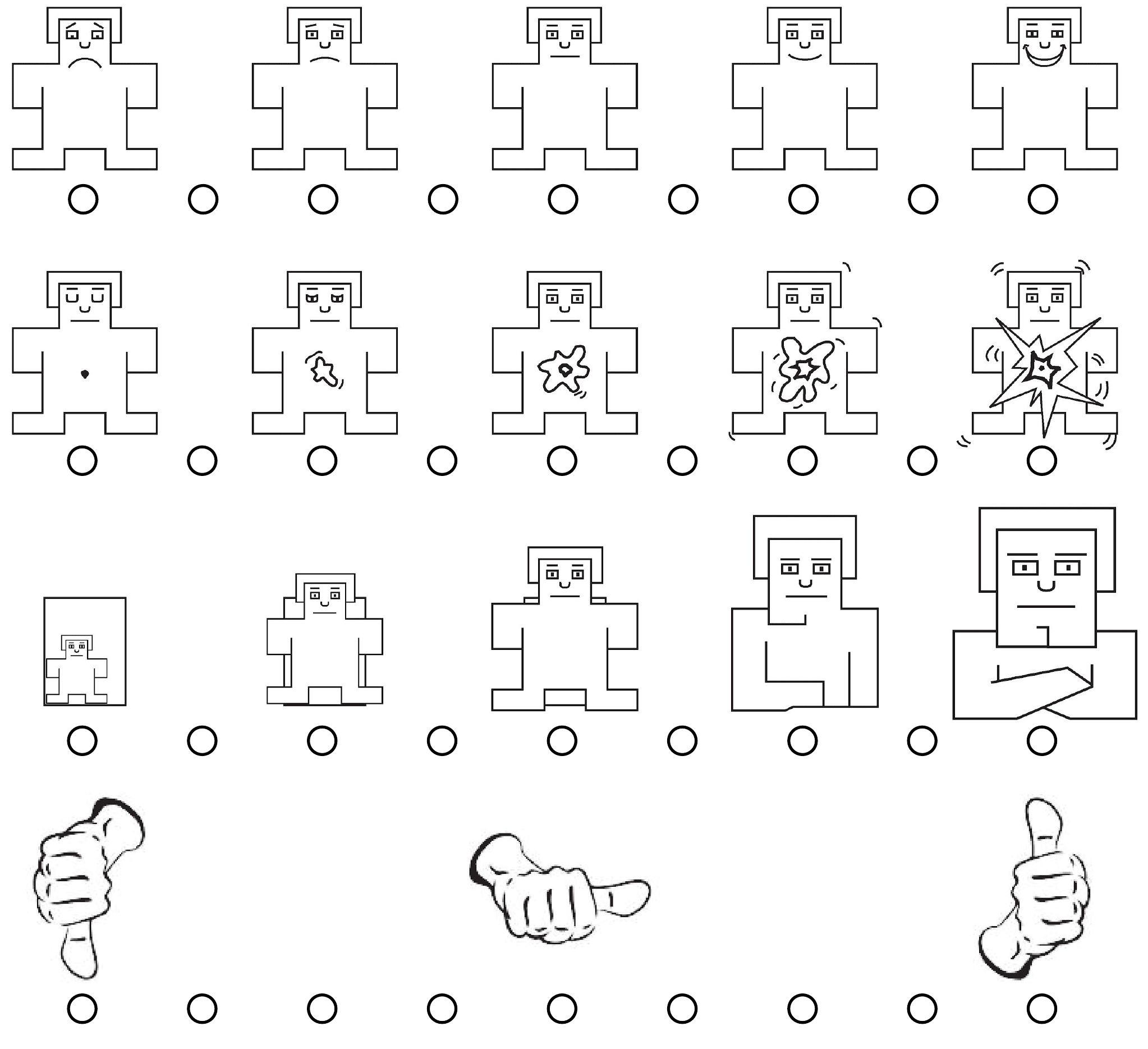}
    \caption{From top down, the pleasure, arousal, dominance, and liking dimensions visualized by means of the extended version of SAM by Koelstra \etal with nine-point rating scales to self-assess each dimension~\cite{Koelstra:2012}.}\label{fig:SAM}
\end{figure}

\section{Background and Related Work}\label{sec:relatedWork}
In this section, we report background information and work investigating developers' affective states. We also provide evidence on the effects of \tdd. 

\subsection{Affective States and Studies about Developers' Affective States}
In psychology, affective states are due to a set of stimuli and directed toward such stimuli. 
They can be characterized according to two theories, discrete and dimensional~\cite{Rus03}.
The former states that there is a fixed set that can be firmly distinguished (\eg resulting in joy, fear, or disgust). The latter characterizes affective states over three orthogonal dimensions: pleasure, arousal, and dominance. Pleasure varies from unpleasant (\eg sad/unhappy) to pleasant (\eg joyful/happy). Arousal varies from inactive (\eg calm/bored) to active (\eg stimulated/excited). Finally, dominance ranges from a helpless and weak feeling (\ie \textit{``without control''}) to an empowered one (\ie \textit{``in control''})~\cite{Koelstra:2012}. 


SAM is a non-verbal self-assessment method for a person's affective reaction based on the dimensional theory and it is used to measure pleasure, arousal, and dominance associated with a stimulus~\cite{Bradley:1994}. Each dimension is described graphically and evaluated thanks to a rating scale---usually a nine-point rating scale---placed below the graphical representation of that dimension (Figure~\ref{fig:SAM}). 
For example, pleasure is visualized by means of figures ranging from an unhappy figure to a happy one.
SAM was extended by Koelstra \etal~\cite{Koelstra:2012}, who added the liking dimension. This dimension ranges from dislike to like and is visualized through thumb-down, -middle, and -up symbols with the rating scale placed below these symbols (Figure~\ref{fig:SAM}).    
SAM is used in Human-Computer Interaction (HCI) and affective computing studies~\cite{HPM05,Koelstra:2012,RMZ03}; lately Software Engineering (SE) work has used this method to study developers' affective states~\cite{GLN18,GWA13}. 

Graziotin~\etal~\cite{GWA13} showed that happier developers are more productive. 
They studied eight developers working on individual projects. 
Every ten minutes, they measured the developers' affective states using SAM and their productivity using a self-assessment questionnaire. 
The results of a mixed-effect model show that pleasure, arousal, and dominance explained 25\% of the variance in productivity.
A follow-up multi-method study with 317 professional developers~\cite{GFW18} showed that both happiness and unhappiness are experienced in relation to increased and decreased productivity and quality of the development process.
A survey of 49 developers provides further evidence that affective states influence the productivity of software developers~\cite{Wro13}. 
In particular, positive ones enhance development productivity, whereas negative ones---particularly frustration---are associated with decreased productivity. 
In an interview with 45 professional developers, Ford and Parnin~\cite{FP15} showed that frustration can occur due to the difficulty of constructing a mental model of the code, learning new tools, dealing with too large task sizes, on boarding a new project, accurate effort estimation, dealing with teammates. 
Mueller and Fritz~\cite{MF15} investigated frustration---and its counterpart, progress or flow~\cite{Csi97}---using biometrics. 
Physiological signals are suited to distinguish the affective states experienced by software developers.
The authors studied 17 novice developers, equipped with three biometric sensors, performing software evolution.
Their results show that different affective states are correlated with the perceived (\ie self-assessed) progress. 

Developers' affective states can be identified in the textual artifact produced during software development (\eg~commit messages). Murgia~\etal~\cite{MTA14} analyzed 17 open-source projects to investigate whether and to what extent issue reports contain information that can be related to specific affective conditions.
They showed that developers express mostly positive affects.
Mantyla~\etal~\cite{MAD16} investigated the association between developers' affective states and productivity by applying sentiment analysis to 700,000 Jira issue reports.
The authors showed that different pleasure is associated with different types of issues (\eg enhancement vs. bug fix request).

Only a few studies assessed affective reactions of developers while performing a task in a controlled fashion.
An example is the work of Khan~\etal~\cite{KBH11}.
The authors linked the effect of mood on debugging in two experiments.
In the first, they elicited specific affective states of 72 developers, who then performed debugging. 
The results show a significant difference in performance between the developers exposed to a stimulus eliciting low arousal and the ones exposed to low arousal. 
In the second, 19 developers worked on a debugging task for 16 minutes, then performed physical exercise, and finally continued working on that  task. 
After the physical exercise, the authors reported increased arousal and pleasure correlated with better task performance.

\subsection{Effects of \tdd}
The effects of TDD on a number of outcomes (\eg developers' productivity) is the subject of several empirical studies, summarized in Systematic Reviews (SR) and Meta-Analysis (MA).
Turhan \etal's SR~\cite{BLD10} includes 32 primary studies (\eg case studies) investigating \tdd in different settings (\eg industry and academia). 
The results are inconsistent, as they show a positive effect on quality, but not regarding productivity.
Rafique and Misic~\cite{RM13} conducted an MA of 25 controlled experiments published between 2000 and 2011. 
Overall, the results are mixed. 
However, \tdd seems to improve quality to the cost of a loss in productivity when considering subjects from academia. 
Finally, Munir \etal's SR~\cite{MMP14} took into account 41 primary studies.
The results show, for both student and professional developers, that \tdd increases quality but not productivity. 

\section{Experiment Planning}\label{sec:planning}
To conduct our experiment, we followed Wohlin \etal's guidelines~\cite{Wohlin:2012}. We report the planning of this experiment based on Jedlitschka \etal's template~\cite{Jedlitschka:2008}.

\subsection{Goals}
We studied the following Research Question (RQ): 
\begin{description}[leftmargin=*]
\item[RQ1.] Is there a difference in the affective reactions of novice developers to a development approach
(\ie \tdd vs. a non-\tdd one)?
\end{description}
With RQ1, we aimed to understand the affective reactions of novice developers due to the use of \tdd in terms of pleasure, arousal, dominance, and liking.
A positive (or negative) effect of \tdd with respect to these four dimensions might imply that \tdd developers are more (or less) effective when performing development tasks. We deepened our investigation by focusing on two central phases of the process underlying \tdd: testing and implementation.\footnote{Although refactoring is part of the process underlying \tdd, we did not consider this phase because refactoring could not be performed when following a non-TDD development approach (and some participants who used a non-TDD approach did not refactor their code).} To this end, we considered the effect of \tdd in terms of the four above-mentioned dimensions when testing and implementing code. Accordingly, we devised two further RQs:
\begin{description}[leftmargin=*]
\item[RQ2.] Is there a difference in the affective reactions of novice developers to the implementation phase when comparing \tdd to a non-\tdd development approach?
\item[RQ3.] Is there a difference in the affective reactions of novice developers to the testing phase when comparing \tdd to a non-\tdd development approach?
\end{description}


\subsection{Experimental Units}
The participants of the experiment were 29 final-year undergraduate students in Computer Science (CS) at the University of Basilicata.
In particular, the students were enrolled in the SE course, which represents the context of our experiment. 
To encourage participation in the study, we informed the students that, regardless of the outcomes they would achieve in the experiment, they would be rewarded with two bonus points on the course final mark.
We can consider final-year undergraduates in CS as a proxy of novice software developers~\cite{Host:2000,SMJ15}. 

Before the SE course, the participants had passed exams related to Procedural and Object Oriented Programming.
During these courses, all students had acquired programming experience in C and Java. According to the curricula, the students did not have a notion of \tdd. 
We also verified that they had never practiced \tdd.  
We trained the participants with a series of both frontal and laboratory lessons after which they performed three homework assignments (\ie development tasks) in preparation for the experiment. 
The lessons covered unit testing, JUnit, Test-Last (TL) development,\footnote{In TL development, a developer first implements a feature entirely and then tests it.} Incremental Test-Last (ITL) development,\footnote{In ITL development, a developer alternates implementing a code increment with testing that increment until the entire feature is implemented.} and TDD. Initially, 47 students accepted to take part in the experiment; 29 completed the training. This sample is homogeneous in terms of skills because of the training process the students underwent (Section~\ref{sec:procedure}), their similar academic background, and their similar self-reported programming experience related to their classmates.\footnote{Over a five-point rating scale---in which the higher the score, the better it is---: MIN=2, MEDIAN=3, MAX=4, and MAD=1.}    


\subsection{Experimental Material}
The experimental objects consisted of the specifications of two development tasks to be implemented in the Java programming language: Bowling Score Keeper (BSK)---an API for calculating the score of a bowling game including bonus---and Mars Rover API (MRA)---an API for controlling the movements of a rover on a 2D planet on which obstacles are present.
Regardless of the experimental object, we provided  the students with the following experimental material: \textit{(i)}~a brief description of the program (\ie a problem statement); \textit{(ii)}~a series of features to implement reported as a set of user stories; \textit{(iii)}~a template project for the Eclipse IDE containing stubs of the expected API signatures and an example JUnit test class; and \textit{(iv)}~an acceptance test suite, developed by the authors, to simulate customers' acceptance of the user stories. 
The acceptance tests were executed using the Concordion framework.\footnote{\url{https://concordion.org/}} We opted for \bsk and \mra as experimental objects because they are often adopted to learn/practice \tdd and were used in past empirical studies on \tdd~\cite{EMT05,Fucci:2018:ESEM,Fucci:2016,SMJ15}.

To gather the affective reactions, we relied on the extended version of SAM by Koelstra \etal~\cite{Koelstra:2012}, which includes four dimensions: pleasure, arousal, dominance, and liking. Each dimension was thus measured through a nine-point rating scale. 


\subsection{Tasks} \label{sec:tasks}
We asked the participants to carry out one development task each, in which they tackled either \bsk or \mra. 
That is, we asked them to implement the user stories associated with these programs---\mra had 11 user stories, while \bsk had 13 user stories---by following \tdd or an alternative approach. 
The participants were asked to take into account one user story at a time (starting from the first one). 
The participant could implement the next user story only when the current one passed its related acceptance test suite. The total time allotted to accomplish the task was three hours.
Right after the development task, we asked the participants to self-assess their affective reactions---in terms of pleasure, arousal, dominance, and liking---of the development approach using SAM. 
Similarly, they self-assessed their affective reactions to the testing and implementation phases. 

\begin{table*}[t]
\centering
\tablescript
\caption{Summary of the dependent variables.}
\label{tab:summaryVars}
\begin{tabular}{lll}
\toprule
 {Name} & {Values} & {Description} \\ \midrule
 \apppls & 1--9 & Affective reaction to the development approach in terms of pleasure. \\
              \appars & 1--9 & Affective reaction to the development approach in terms of arousal.           \\
              \appdom & 1--9 & Affective reaction to the development approach in terms of dominance.             \\
               \applik & 1--9 & Affective reaction to the  development approach in terms of liking.            \\ \addlinespace
              \imppls & 1--9 & Affective reaction to the implementation phase in terms of pleasure.               \\
              \impars & 1--9 & Affective reaction to the implementation phase in terms of arousal.           \\
              \impdom & 1--9 & Affective reaction to the implementation phase in terms of dominance.            \\
              \implik & 1--9 & Affective reaction to the implementation phase in terms of liking.          \\ \addlinespace
              \tespls & 1--9 & Affective reaction to the testing phase in terms of pleasure.            \\
               \tesars & 1--9 & Affective reaction to the testing phase in terms of arousal.            \\
              \tesdom & 1--9 & Affective reaction to the testing phase in terms of dominance.            \\
               \teslik & 1--9 &  Affective reaction to the testing phase in terms of liking.  \\ \bottomrule
\end{tabular}
\end{table*}

\subsection{Hypotheses, Parameters, and Variables}
We manipulated two independent variables: \textit{Approach} and \textit{Object}. 
The former represents the development approach the participants had to follow to carry out the development task, namely \tdd or the approach they preferred (\ie \yw). Therefore, Approach is a categorical variable with two values, \tdd and \yw.
The Object variable indicates the experimental object the participants dealt with (\ie \bsk or \mra) in the experiment. 
Similarly to Approach, Object is a categorical variable. It can assume the following two values: \bsk and \mra.

To measure PLeaSure (\pls), ARouSal (\ars), DOMinance (\dom), and LIKing (\lik) associated with the development APProach (APP), we used the following ordinal dependent variables: 
\texttt{\apppls}, \texttt{\appars}, \texttt{\appdom}, and \texttt{\applik}.
Similarly, we quantified pleasure, arousal, dominance, and liking for the IMPlementation (IMP) and TESting (TES) phases by means of the following ordinal dependent variables: \texttt{\imppls}, \texttt{\impars}, \texttt{\impdom}, \texttt{\implik}, \texttt{\tespls}, \texttt{\tesars}, \texttt{\tesdom}, and \texttt{\teslik}.
In Table~\ref{tab:summaryVars}, we summarize the dependent variables of our experiment.

We formulated and tested the following null hypotheses:
\begin{description}[leftmargin=*]
\item[$\boldsymbol{H0_X}$.] There is no difference between \tdd and \yw with respect to the dependent variable $X \in $ \{\texttt{\apppls}, \texttt{\appars}, \texttt{\appdom}, \texttt{\applik}, \texttt{\imppls}, \texttt{\impars}, \texttt{\impdom}, \texttt{\implik}, \texttt{\tespls}, \texttt{\tesars}, \texttt{\tesdom}, \texttt{\teslik}\}.
\end{description}

\subsection{Experiment Design}
The design of our experiment was 2x2 factorial---a type of between-subjects design~\cite{Wohlin:2012}. 
In particular, each participant used only one development approach (\ie either \tdd or \yw).
Within each development approach, each participant tackled only one experimental object---\ie either \bsk or \mra. 
Those who used \tdd (either tackling \bsk or \mra) form the treatment group, while those who experimented \yw (either tackling \bsk or \mra) form the control group. 

In Table~\ref{tab:design}, we show the number of participants assigned  to each of four groups constituted by the combination of development approaches and experimental objects. The assignment was randomly performed.
By looking at Table~\ref{tab:design}, we can notice that the number of participants distributed among development approaches, experimental objects, and their combination was almost uniform.  

\begin{table}[t]
\centering
\tablescript
\caption{Number of participants assigned to each studied approach and object.}
\label{tab:design}
\begin{tabular}{llll}
\toprule
 &  & \multicolumn{2}{l}{Approach} \\ \cmidrule{3-4} 
 &  & TDD & YW \\ \midrule
{Object} & {MRA} & 7 & 7 \\
 & {BSK} & 8 & 7 \\ \bottomrule
\end{tabular}
\end{table}
\subsection{Procedure}\label{sec:procedure}
The experimental procedure included the following steps.  
\begin{enumerate}[leftmargin=*]
    \item We gathered the availability of the students to participate in the experiment through a questionnaire (also used to gather demographic information). 
    
    \item The participants attended the frontal lessons on unit testing, JUnit, TL development, and ITL development. They also took part in a laboratory session (of two hours) on unit testing with JUnit.
    
    \item We (randomly) split the participants into two groups: \tdd and \yw. The participants in the \yw and \tdd groups were 14 and 15, respectively (Table~\ref{tab:design}). Based on the group, the participants underwent two different training:
    \begin{itemize}[leftmargin=*]
        \item The students in the \tdd group attended a face-to-face lesson  on \tdd and experimented this approach through two laboratory sessions (of two hours each) and three homework assignments. Handing in the assignments was mandatory to participate in the experimental session.
        \item The students in the \yw group did not attend lessons on \tdd nor used the approach in the laboratory sessions and assignments. However, the students in the \yw group took part in two laboratory sessions (of two hours each) and performed the same homework assignments as the \tdd group, but to practice TL and ITL. Similarly to the \tdd group, homework assignments were mandatory.
    \end{itemize}
    
    \item  The experimental session took place under controlled conditions in a research laboratory at the University of Basilicata.
    All the laboratory computers were equipped with the same hardware and software. Furthermore, they contained all the material necessary to complete the tasks, \ie the template project (of Eclipse) corresponding to the assigned experimental object. During the experimental session, the participates performed the development tasks and then they self-assessed their affective reactions (Section~\ref{sec:tasks}). We avoided interactions among participants by monitoring them during the task execution. 
    

    
    
\end{enumerate}

\subsection{Analysis Procedure}
We relied on diverging stacked bar plots to summarize the distributions of the values of the dependent variables. 
To test the null hypotheses (one for each dependent variable), we used  a non-parametric version of ANOVA, namely ANOVA Type Statistic (ATS)~\cite{Brunner:1997}. We opted for ATS because this method is frequently used in the medical field and recommend, in place of ANOVA, in the HCI field to analyze data from rating scales in factorial designs like ours~\cite{Kaptein:2010}. 
For each dependent variable X, we built ATS models as follows: 
 $   X \sim Approach + Object + Approach:Object$.
Approach and Object are the variables we manipulated, while Approach:Object represents their interaction. 
That is, this model allows determining if Approach, Object, and Approach:Object had statistically significant effects on a given dependent variable.
To judge whether an effect is statistically significant, we used $\alpha=0.05$ as the threshold value. It indicates 5\% chance that a Type-I-error occurs (\ie rejecting the null hypothesis when it is true)~\cite{Wohlin:2012}. If a p-value is less than $\alpha$, it is deemed statistically significant.
In case of a statistically significant effect of Approach, we quantified the magnitude of that effect through the Cliff's $\delta$ effect size. We opted for such a kind of effect size since it was originally developed for use with ordinal variables (like ours)~\cite{Cliff:1996}. The effect size is considered: \textit{negligible} if $|\delta|<$ 0.147, \textit{small} if $0.147\le|\delta|<$ 0.33, \textit{medium} if $0.33\le|\delta|<$ 0.474, or \textit{large} if $|\delta|\ge$ 0.474~\cite{Romano:2006}.

\textbf{Further Analysis.}
To better contextualize our experiment, we also assessed participants' performance. 
We counted the number of user stories each participant implemented in the allotted time.
We normalized them in the [0, 1] interval to obtain a fair comparison between participants tackling tasks with a different number of user stories.
We named this additional dependent variable \texttt{\str}. 
The strategy we followed to quantify participants' performance is \textit{time-fixed}---the number of successful steps within a fixed time span defines performance~\cite{Bergersen:2014}. 
The higher the value of \str, the better the developer's performance.

\section{Results}\label{sec:results}

\begin{figure}[t]
   \centering
    \includegraphics[width=\textwidth]{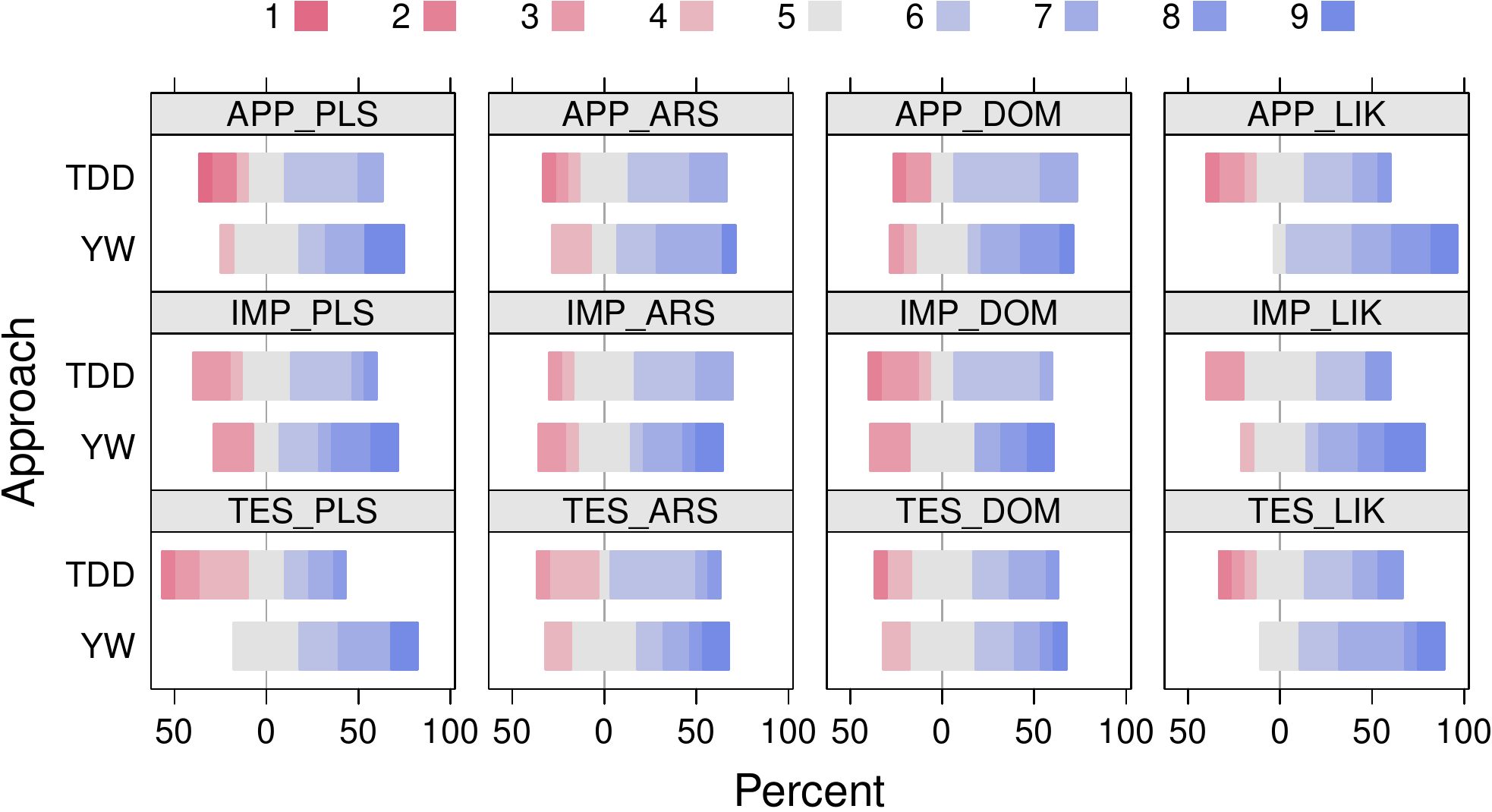}
    \caption{Diverging stacked bar plots for the dependent variables.}\label{fig:plot}
\end{figure}

In Figure~\ref{fig:plot}, we show the diverging stacked bar plots summarizing the distributions of the values of the twelve dependent variables. The x-axes report the frequencies of the dependent variable values, which range from one---the most negative value---to nine---the most positive value. Therefore, the neutral value is five. The diverging stacked bar plots display positive values in shades of blue, while those negative in shades of red. The neutral value is displayed in grey. The y-axes allow grouping the values based on the Approach variable. As for the results from ATS, they are summarized in Table~\ref{tab:ats}

\begin{table}[t]
\tablescript
\centering
\caption{Results from ATS---F-statistic (in parentheses) and p-values (in bold those less than $\alpha=0.05$) for the dimensions associated with the development approach, and implementation and testing phases.}\label{tab:ats}
\begin{tabular}{llll}\toprule
Dep. Var.                    & \multicolumn{3}{l}{Indep. Var.} \\ \cmidrule{2-4}
                  & Approach & Object & Approach:Object \\ \midrule
\apppls & 0.1615 (2.1094) & 0.7721 (0.0861)  & 0.8998 (0.0162) \\
\appars & 0.2774 (1.2378) &  0.7794 (0.0803) & 0.1816 (1.8985) \\
\appdom & 0.2796 (1.2313) & 0.8569 (0.0333)  & 0.4296 (0.6487) \\
\applik & \textbf{0.0024} (11.4580) & 0.1650 (2.0467) & 0.6368 (0.2285) \\ \addlinespace

\imppls & 0.2008 (1.7454) & 0.6663 (0.1914) & 0.9793 (0.0007) \\
\impars & 0.6799 (0.1755) & 0.6881 (0.1661) & 0.5752 (0.3249) \\
\impdom & 0.3449 (0.9330)  &  0.5614 (0.3480) &  0.4672 (0.5481) \\
\implik & \textbf{0.0396} (4.7562)  &  0.1862 (1.8557)  &  0.2703 (1.2752) \\ \addlinespace

\tespls & \textbf{0.0178} (6.5782) & 0.6500 (0.2118)  & 0.7652 (0.0915)  \\
\tesars & 0.4147 (0.6887) & 0.4765 (0.5230)  & 0.3406 (0.9451) \\
\tesdom & 0.6341 (0.2324) &  0.2564 (1.3508) & 0.4738 (0.5293) \\
\teslik & 0.0504 (4.2785) & 0.1194 (2.6224) & 0.0547 (4.1112) \\ 
\bottomrule
\end{tabular}
\end{table}


    
    

\textbf{RQ1---Affective Reactions to Development Approach.} By looking at Figure~\ref{fig:plot}, there is no noticeable difference between \tdd and \yw regarding pleasure (\apppls), arousal (\appars), and dominance (\appdom). However, we can notice a slight trend in favor of \yw since \tdd tends to appear with some frequency towards very negative scores (\ie $<$ 4) more than \yw. As for liking (\applik), Figure~\ref{fig:plot} suggests that participants in the \yw group liked this approach more, compared to the participants in the \tdd group.  




The ATS results (Table~\ref{tab:ats}) indicate that there is no statistically significant difference between \tdd and \yw regarding pleasure, arousal, and dominance. 
Accordingly, we cannot reject the corresponding null hypotheses. 
The test results allow us to reject $H0_\applik$, showing an effect of the development approach on \applik. 
The frequencies displayed in Figure~\ref{fig:plot} suggest that such an effect is in favor of \yw. The effect size is \textit{large} ($\delta=$ 0.6048, CI95\% = [0.2018, 0.8326]).

Based on these results, we can answer RQ1 as follows: 
\textit{developers using TDD seem to like their development approach less than those using a non-\tdd one.}


    
    

\textbf{RQ2---Affective Reactions to Implementation Phase.}
Figure~\ref{fig:plot} does not highlight remarkable difference between \tdd and \yw for pleasure (\imppls), arousal (\impars), and dominance (\impdom) during the implementation phase. However, for these dimensions, we can observe a slight trend in favor of \yw since the percentages of very positive scores (\ie $>$ 6) appear to be higher for \yw. 
With respect to the liking dimension (\implik), Figure~\ref{fig:plot} suggest that participants who followed \yw liked the implementation phase more, compared to the ones following \tdd. 



The results in Table~\ref{tab:ats} do not show a statistically significant difference between \tdd and \yw regarding pleasure, arousal, and dominance. 
Accordingly, we cannot reject the null hypotheses corresponding to these dimensions.
We reject $H0_\implik$ as there is a statistically significant effect of Approach on \implik. 
The effect is in favor of \yw as the plot in Figure~\ref{fig:plot} suggest. 
The size of the effect of Approach is \textit{medium} ($\delta=$ 0.4286, CI95\% = [0.0209, 0.714]).

According to the obtained results, we can answer RQ2 as follows: \textit{developers using TDD seem to like the implementation phase less than those using a non-\tdd development approach.}

\textbf{RQ3---Affective Reactions to Testing Phase.}
Figure~\ref{fig:plot} suggests that there is a difference between \tdd and \yw in terms of pleasure (\tespls) during  the  testing  phase. In particular, the participants using \tdd reported negative scores with some frequency while those using \yw never reported negative scores. 
When considering the arousal (\tesars) and dominance (\tesdom) dimensions, we cannot observe any substantial difference between the two development approaches  (Figure~\ref{fig:plot}). On the contrary, when considering liking (\teslik), we can notice a difference between \tdd and \yw in favor of the latter as \yw tends to have more very positive scores (\ie $>$ 6) than \tdd. 




The results of ATS (Table~\ref{tab:ats}) reveal a statistically significant difference for the pleasure dimension, which allows us to reject the $H0_{\tespls}$ hypothesis.  
Such a difference is in favor of \yw (Figure~\ref{fig:plot}).
The effect size is \textit{large} ($\delta=$ 0.5, CI95\% = [0.0796, 0.7694]).
As for arousal and dominance, the effect of the development approach is not statistically significant during the testing phase. Regarding liking, the observed difference in \teslik between \yw and \tdd is not significant. 

The obtained results allowed us to answer RQ3 as follows: \textit{the testing phase seems to make developers using TDD less happy compared to those using a non-TDD development approach.}


\textbf{Further Analysis Results.}\label{sec:furtherAnalysis}
We also studied participants' performance by running ATS using \str as dependent variable.\footnote{\str does not meet the normality assumption (Shapiro-Wilk normality test p-value $=$ 0.0114); this is why we run ATS (rather than ANOVA).} 
The results indicates that Approach (p-value $=$ 0.4765), Object (p-value $=$ 0.2596) and their interaction (p-value $=$ 0.0604) have no statistically significant effect on \str.

\section{Discussion}\label{sec:discussion}

The results from this experiment present initial evidence about aspects that are not investigated by the empirical TDD research. 
Current research on the effects of TDD shows inconclusive results~\cite{MMP14,RM13,BLD10}, which can be attributed to the disliking the developers experience when using TDD, at least in the experiment time frame. 
We show initial evidence---supported by a large effect size---that, although participants' performance do not vary significantly (Section~\ref{sec:furtherAnalysis}) due to the development approach, \tdd seems to negatively impact  affective reactions (\ie liking) of novice developers. 
Researchers need to be aware of the effect that disliking \tdd can have (\eg low motivation to perform a task) when designing experiments involving such an approach.

We observed a difference between \tdd and \yw regarding the liking dimension for the implementation phase.
The medium effect size shows initial evidence that implementing production code when performing \tdd seems to be disliked by developers. 
Writing production code during \tdd is trivial, at least in the first few iterations, and usually consists in taking shortcuts (\eg returning hard-coded values) to make the test pass. 
In our study, developers did not like such an activity.
We conjecture this may be the case because they did not base their implementation on creative activities requiring challenging decisions. 
Conversely, this should have resulted in different levels of arousal (\ie low for \tdd) compared to non-TDD developers which we did not observe. 
Our explanation for the lack of such an observation lies in the task complexity which could have not been enough to elicit stronger arousal responses.
The lack of significant effect due to the Object in our ATS models partially supports this explanation. 

The liking dimension could change over time. Longitudinal studies could be necessary to validate such hypothesis and qualitative studies are required to pinpoint the reason for the observed results. 
In particular, the latter is necessary to explain the contrasting interview results presented in Romano~\etal~\cite{RFS17} in which a preference for the implementation phase among TDD developers emerged due to its rewarding feeling (\ie observing the JUnit red bar turn green).


The testing phase seems to make developers using \tdd less happy than those using a non-TDD approach.
Previous work~\cite{RFS17} shows that \tdd developers create a mental model of their solution to a task which is then translated into unit tests. 
Novice developers can be uncomfortable with such an activity due to the counter-intuitiveness of this step, but also due to the difficulty of writing tests of good granularity in the absence of the underlying production code~\cite{FET17,KT18}.
Conversely, developers following the non-TDD approach can decide when and what to test without (mindlessly) following a process.
Such freedom of action---\eg testing what is worth according to the developer's own understanding---can explain the higher pleasure score of non-TDD developers. 
Although this can be the case in the short term, longitudinal studies of 
\tdd developers' affective states are also necessary in this case.

In general, our observations are supported by the results of a survey among professional developers, who are new to TDD~\cite{AFG11}.
They expressed concerns that worrying about writing unit tests and working in small increments distracts them from achieving their implementation goals while the extra effort necessary to perform TDD is perceived as waste~\cite{AFG11}. 
Practitioners should take into account the results of this study when introducing TDD. 
The disliking attitude towards this development approach can (negatively) impact developers' performance in the long run (which we did not observe in the short term).
Considering the results regarding the (negative) affective reactions to the implementation and testing phases, we suggest that, for greenfield development tasks, developers could skip TDD for few initial iterations and rely on their preferred development approach.
This should not have an impact on performance but could reduce their negative affect which, in turn, could impact motivation and job satisfaction~\cite{GFW18,BLD10}.

\section{Threats to Validity}\label{sec:threats}
We discuss the threats that could affect the validity of the results according to the guidelines presented by Wohlin~\etal~\cite{Wohlin:2012}. 
We ranked these threats from the most to the least sensible for the goal of our study. 
In particular, being this the first investigation of developers' effective states when using TDD, we prioritize threats to internal validity. 
That is, we were more interested in studying that cause-effect relationships were correctly identified.

\textbf{Internal Validity.} 
A possible threat is the voluntary participation in the study (\ie \textit{selection threat)} by students particularly willing to be assessed. However, we limited this threat by embedding the experiment in the SE course and did not consider its outcome when grading.
To deal with a \textit{threat of diffusion or treatments imitations}, two authors of this paper monitored participants to prevent them from exchanging information during the experiment.
Another threat might be \textit{resentful demoralization}---participants assigned to a less desirable treatment might not perform as good as they normally would.

\textbf{Construct Validity.}
Each dependent variable was measured by means of a single self-assessment at the end of the task. If there was a measurement bias, the results would be misleading (\ie \textit{mono-method bias threat}). 
Although the participants were not informed about the research goals of our experiment, they might guess them and change their behavior accordingly (\ie \textit{threat of hypotheses guessing}). To deal with an \textit{evaluation apprehension threat}, we did not evaluate the participants in the experiment on the basis of their performances. 
We acknowledge the presence of a \textit{threat of restricted generalizability across constructs}. That is, while influencing the affective states, the approach might affect other non-measured constructs (\eg cognitive load). 

\textbf{Conclusion Validity.}
To mitigate a \textit{threat of random heterogeneity of participants}, our sample included students who followed the same course at the same university, underwent a similar training, and had similar background, skills and experience. 
A \textit{threat of reliability of treatment implementation} might occur (\eg some participants might follow \tdd more strictly than others so influencing their affective reactions). 
In several occasions, during the task execution, we reminded the participants to follow the treatment they were assigned to.
Finally, our sample was limited because of the difficulty of recruiting participants available for all the period of the experiment including training.

\textbf{External Validity.}
The participants in our study were undergraduate students. This could pose some threats to the generalizability of the  results to the population of professional developers (\ie \textit{threat of interaction of selection and treatment}). However, the use of students has the advantage that they have homogeneous background and are particularly suitable to obtain preliminary evidence from empirical studies~\cite{Carver:2003}. Therefore, the use of students could be considered appropriate, as suggested in the literature~\cite{Carver:2003,Host:2000}. 
The used experimental objects might pose a \textit{threat of interaction of setting and treatment}. 
\bsk and \mra can be completed in a single exercise session of three hours~\cite{Fucci:2018:ESEM,Fucci:2016} so allowing a better control over the participants. This was our preferred trade-off due to the theory-testing nature of our experiment.

\section{Conclusions}\label{sec:conclusion}
We presented a controlled experiment to study whether and in what phase \tdd influences affective states of novice developers in terms of pleasure, arousal, dominance, and liking. 
Developers in the treatment group implemented a task using \tdd whereas the control group used a non-\tdd development approach (\ie \yw).  
We compared the affective reactions of developers with respect to the development approach they used, further focusing on the implementation and the testing phases. 
The results indicate a significant difference between the two development approaches  in terms of affective reactions. Developers seem to like   \yw   more than \tdd. Moreover, developers like the implementation phase in \yw more than that in \tdd and the testing phase makes developers using \tdd less happy. The findings from our study can help explain the inconclusive results of experiments focusing on the claimed effect of TDD. 
As future work, 
we plan to conduct replications, investigations focusing on settings closer to the real world, and longitudinal studies to measure affective states in the long run.


%
%
%

\bibliographystyle{splncs04}
\bibliography{bibliography}

\begin{thebibliography}{10}
\providecommand{\url}[1]{\texttt{#1}}
\providecommand{\urlprefix}{URL }
\providecommand{\doi}[1]{https://doi.org/#1}

\bibitem{AFG11}
Aniche, M.F., Ferreira, T.M., Gerosa, M.A.: What concerns beginner test-driven
  development practitioners: a qualitative analysis of opinions in an agile
  conference. In: Proceedings of Brazilian Workshop on Agile Methods (2011)

\bibitem{Bec03}
Beck, K.: Test-driven development: by example. Addison-Wesley (2003)

\bibitem{Bergersen:2014}
{Bergersen}, G.R., {Sj{\o}berg}, D.I.K., Dyb{\aa}, T.: Construction and
  validation of an instrument for measuring programming skill. IEEE
  Transactions on Software Engineering  \textbf{40}(12),  1163--1184 (2014)

\bibitem{Bradley:1994}
Bradley, M.M., Lang, P.J.: Measuring emotion: The self-assessment manikin and
  the semantic differential. Journal of Behavior Therapy and Experimental
  Psychiatry  \textbf{25}(1),  49 -- 59 (1994)

\bibitem{Brunner:1997}
Brunner, E., Dette, H., Munk, A.: Box-type approximations in nonparametric
  factorial designs. Journal of the American Statistical Association
  \textbf{92}(440),  1494--1502 (1997)

\bibitem{Carver:2003}
Carver, J., Jaccheri, L., Morasca, S., Shull, F.: Issues in using students in
  empirical studies in software engineering education. In: Proceedings of
  International Symposium on Software Metrics. pp. 239--249. IEEE (2003)

\bibitem{Cliff:1996}
Cliff, N.: Ordinal Methods for Behavioral Data Analysis. Erlbaum (1996)

\bibitem{CHG10}
Colomo-Palacios, R., Hern{\'a}ndez-L{\'o}pez, A., Garc{\'\i}a-Crespo, {\'A}.,
  Soto-Acosta, P.: A study of emotions in requirements engineering. In:
  Proceedings of World Summit on Knowledge Society. pp.~1--7. Springer (2010)

\bibitem{Csi97}
Csikszentmihalyi, M.: Finding flow: The psychology of engagement with everyday
  life. Basic Books (1997)

\bibitem{EMT05}
Erdogmus, H., Morisio, M., Torchiano, M.: On the effectiveness of the
  test-first approach to programming. {IEEE} Trans. Software Eng.
  \textbf{31}(3),  226--237 (2005)

\bibitem{FP15}
Ford, D., Parnin, C.: Exploring causes of frustration for software developers.
  In: Proceedings of International Workshop on Cooperative and Human Aspects of
  Software Engineering. pp. 115--116. IEEE Press (2015)

\bibitem{FET17}
Fucci, D., Erdogmus, H., Turhan, B., Oivo, M., Juristo, N.: A dissection of the
  test-driven development process: does it really matter to test-first or to
  test-last? IEEE Transactions on Software Engineering  \textbf{43}(7),
  597--614 (2017)

\bibitem{Fucci:2018:ESEM}
Fucci, D., Romano, S., Baldassarre, M.T., Caivano, D., Scanniello, G., Turhan,
  B., Juristo, N.: A longitudinal cohort study on the retainment of test-driven
  development. In: Proceedings of International Symposium on Empirical Software
  Engineering and Measurement. pp. 18:1--18:10. ESEM '18, ACM (2018)

\bibitem{Fucci:2016}
Fucci, D., Scanniello, G., Romano, S., Shepperd, M., Sigweni, B., Uyaguari, F.,
  Turhan, B., Juristo, N., Oivo, M.: An external replication on the effects of
  test-driven development using a multi-site blind analysis approach. In:
  Proceedings of International Symposium on Empirical Software Engineering and
  Measurement. pp. 3:1--3:10. ESEM '16, ACM (2016)

\bibitem{GLN17}
Gachechiladze, D., Lanubile, F., Novielli, N., Serebrenik, A.: Anger and its
  direction in collaborative software development. In: Proceedings of
  International Conference on Software Engineering: New Ideas and Emerging
  Technologies Results Track. pp. 11--14. IEEE (2017)

\bibitem{GLN18}
Girardi, D., Lanubile, F., Novielli, N., Fucci, D.: Sensing developers’
  emotions: The design of a replicated experiment. In: Proceedings of
  International Workshop on Emotion Awareness in Software Engineering. pp.
  51--54. IEEE (2018)

\bibitem{GFW18}
Graziotin, D., Fagerholm, F., Wang, X., Abrahamsson, P.: What happens when
  software developers are (un) happy. Journal of Systems and Software
  \textbf{140},  32--47 (2018)

\bibitem{GWA13}
Graziotin, D., Wang, X., Abrahamsson, P.: Are happy developers more productive?
  In: Proceedings of International Conference on Product Focused Software
  Process Improvement. pp. 50--64. Springer (2013)

\bibitem{HPM05}
Herbon, A., Peter, C., Markert, L., Van Der~Meer, E., Voskamp, J.: Emotion
  studies in hci-a new approach. In: Proceedings of International Conference on
  Human-Computer Interaction (2005)

\bibitem{Host:2000}
H{\"o}st, M., Regnell, B., Wohlin, C.: Using students as subjects---a
  comparative study of students and professionals in lead-time impact
  assessment. Empirical Software Engineering  \textbf{5}(3),  201--214 (2000)

\bibitem{Jedlitschka:2008}
Jedlitschka, A., Ciolkowski, M., Pfahl, D.: Guide to Advanced Empirical
  Software Engineering, chap. Reporting Experiments in Software Engineering,
  pp. 201--228. Springer London (2008)

\bibitem{JM07}
Jeffries, R., Melnik, G.: Guest editors' introduction: Tdd--the art of fearless
  programming. IEEE Software  \textbf{24}(3),  24--30 (2007)

\bibitem{Kaptein:2010}
Kaptein, M.C., Nass, C., Markopoulos, P.: Powerful and consistent analysis of
  likert-type ratingscales. In: Proceedings of the SIGCHI Conference on Human
  Factors in Computing Systems. pp. 2391--2394. CHI '10, ACM (2010)

\bibitem{KT18}
Karac, I., Turhan, B.: What do we (really) know about test-driven development?
  IEEE Software  \textbf{35}(4),  81--85 (2018)

\bibitem{KBH11}
Khan, I.A., Brinkman, W.P., Hierons, R.M.: Do moods affect programmers’ debug
  performance? Cognition, Technology \& Work  \textbf{13}(4),  245--258 (2011)

\bibitem{Koelstra:2012}
{Koelstra}, S., {Muhl}, C., {Soleymani}, M., {Lee}, J., {Yazdani}, A.,
  {Ebrahimi}, T., {Pun}, T., {Nijholt}, A., {Patras}, I.: Deap: A database for
  emotion analysis using physiological signals. IEEE Transactions on Affective
  Computing  \textbf{3}(1),  18--31 (Jan 2012)

\bibitem{MAD16}
M{\"a}ntyl{\"a}, M., Adams, B., Destefanis, G., Graziotin, D., Ortu, M.: Mining
  valence, arousal, and dominance: possibilities for detecting burnout and
  productivity? In: Proceedings of International Conference on Mining Software
  Repositories. pp. 247--258. ACM (2016)

\bibitem{MF15}
M{\"u}ller, S.C., Fritz, T.: Stuck and frustrated or in flow and happy: sensing
  developers' emotions and progress. In: International Conference on Software
  Engineering. vol.~1, pp. 688--699. IEEE (2015)

\bibitem{MMP14}
Munir, H., Moayyed, M., Petersen, K.: Considering rigor and relevance when
  evaluating test driven development: A systematic review. Information and
  Software Technology  \textbf{56}(4),  375--394 (2014)

\bibitem{MTA14}
Murgia, A., Tourani, P., Adams, B., Ortu, M.: Do developers feel emotions? an
  exploratory analysis of emotions in software artifacts. In: Proceedings of
  working conference on mining software repositories. pp. 262--271. ACM (2014)

\bibitem{OMD16}
Ortu, M., Murgia, A., Destefanis, G., Tourani, P., Tonelli, R., Marchesi, M.,
  Adams, B.: The emotional side of software developers in jira. In: Proceedings
  of the 13th International Conference on Mining Software Repositories. pp.
  480--483. ACM (2016)

\bibitem{RM13}
Rafique, Y., Mi{\v{s}}i{\'c}, V.B.: The effects of test-driven development on
  external quality and productivity: A meta-analysis. IEEE Transactions on
  Software Engineering  \textbf{39}(6),  835--856 (2013)

\bibitem{Romano:2006}
Romano, J., Kromrey, J., Coraggio, J., Skowronek, J.: {Appropriate statistics
  for ordinal level data: Should we really be using t-test and Cohen'sd for
  evaluating group differences on the NSSE and other surveys?} In: Annual
  Meeting of the Florida Association of Institutional Research. pp.~1--3 (2006)

\bibitem{RFS17}
Romano, S., Fucci, D., Scanniello, G., Turhan, B., Juristo, N.: Findings from a
  multi-method study on test-driven development. Information and Software
  Technology  \textbf{89},  64--77 (2017)

\bibitem{RMZ03}
Rudmann, D.S., McConkie, G.W., Zheng, X.S.: Eyetracking in cognitive state
  detection for hci. In: Proceedings of international conference on Multimodal
  interfaces. pp. 159--163. ACM (2003)

\bibitem{Rus03}
Russell, J.A.: Core affect and the psychological construction of emotion.
  Psychological review  \textbf{110}(1), ~145 (2003)

\bibitem{SMJ15}
Salman, I., Misirli, A.T., Juristo, N.: Are students representatives of
  professionals in software engineering experiments? In: International
  Conference on Software Engineering. vol.~1, pp. 666--676. IEEE (2015)

\bibitem{BLD10}
Turhan, B., Layman, L., Diep, M., Erdogmus, H., Shull, F.: How effective is
  test-driven development. In: Making Software: What Really Works, and Why We
  Believe It, pp. 207--217. O'Reilly Media (2010)

\bibitem{Wohlin:2012}
Wohlin, C., Runeson, P., Hst, M., Ohlsson, M.C., Regnell, B., Wessln, A.:
  Experimentation in Software Engineering. Springer Publishing Company (2012)

\bibitem{Wro13}
Wrobel, M.R.: Emotions in the software development process. In: Proceedings of
  International Conference on Human System Interactions. pp. 518--523. IEEE
  (2013)

\end{thebibliography}

\end{document}